\documentclass[11pt,a4paper]{article}
\usepackage{jheppub}

\usepackage{slashed}
\usepackage{youngtab}

\newcommand{\be}{\begin{equation}}
\newcommand{\ee}{\end{equation}}
\newcommand{\ba}{\begin{eqnarray}}
\newcommand{\ea}{\end{eqnarray}}

\newcommand\fverb{\setbox\fverbbox=\hbox\bgroup\verb}
\newcommand\fverbdo{\egroup\medskip\noindent%
            \fbox{\unhbox\fverbbox}\ }
\newcommand\fverbit{\egroup\item[\fbox{\unhbox\fverbbox}]}
\newbox\fverbbox

\newcommand{\nablaslash}{\not{\hbox{\kern-3pt $\nabla$}}}
\newcommand{\nn}{\nonumber}

\title{Thermodynamic and classical instability of AdS black
holes in fourth-order gravity}

\author[a]{Yun~Soo~Myung}
\author[a]{and~Taeyoon~Moon}
\affiliation[a]{Institute of Basic Science and Department of
Computer
Simulation\\ Inje University\\
Gimhae 621-749, Korea}

\emailAdd{ysmyung@inje.ac.kr} \emailAdd{tymoon@inje.ac.kr}

\abstract{We study  thermodynamic and classical instability of AdS
black holes in fourth-order gravity. These include the BTZ black
hole in new massive gravity, Schwarzschild-AdS black hole, and
higher-dimensional AdS black holes in fourth-order gravity.  All
thermodynamic quantities which are computed using the
Abbot-Deser-Tekin method are used to study thermodynamic instability
of AdS black holes. On the other hand, we investigate the $s$-mode
Gregory-Laflamme instability of the massive graviton  propagating
around the AdS black holes.  We establish the connection between the
thermodynamic instability and the GL instability of AdS black holes
in fourth-order gravity. This shows  that the Gubser-Mitra
conjecture holds for AdS black holes found from fourth-order
gravity.}

\begin{document}
\maketitle \flushbottom

\section{Introduction}

Concerning the thermodynamic analysis of a black hole,  the
Schwarzschild black hole in Einstein gravity is in an unstable
equilibrium with the heat reservoir of the temperature
$T$~\cite{GPY}. Its fate under small fluctuations will be either to
decay to  hot flat space by emitting Hawking radiation or to grow
without limit by absorbing thermal radiations in the infinite heat
reservoir~\cite{York}.  This means that an isolated black hole is
never in thermal equilibrium in asymptotically flat spacetimes
because of its negative heat capacity.   Thus,  one has to find  a
way of getting a stable black hole which might be  in an equilibrium
with a finite heat reservoir. A black hole could be rendered
thermodynamically stable by placing it in four-dimensional anti-de
Sitter (AdS$_4$) spacetimes because AdS$_4$ spacetimes play the role
of a confining box.   An important point to understand is to know
how a stable black hole with positive  heat capacity could emerge
from thermal radiation through a phase transition.  The Hawking-Page
(HP) phase transition occurs between thermal AdS spacetimes (TAdS)
and Schwarzschild-AdS (SAdS) black hole~\cite{HP,BCM}, which is
known to be one typical example of the first-order phase transition
in the gravitational system. Its higher dimensional extension and
the AdS/CFT correspondence of confinement-deconfinement phase
transition were studied in~\cite{Witt}.

To study the HP phase transition in Einstein gravity explicitly, we
are necessary  to know the Arnowitt-Deser-Misner (ADM)
mass~\cite{Arnowitt:1962hi}, the Hawking temperature, and the
Bekenstein-Hawking (BH) entropy. These are combined to give the
on-shell Helmholtz free energy in canonical ensemble which
determines the global thermodynamic stability. The other important
quantity is the heat capacity which determines the local
thermodynamic stability. If one uses the Euclidean action approach,
one also finds these quantities consistently~\cite{dbranes}.

However,  the black hole thermodynamics was not completely  known in
fourth-order gravity because one has encountered some difficulty to
compute their conserved quantities in asymptotically AdS spacetimes
exactly.  Recently, there was  some progress on computation scheme
of mass and related thermodynamic quantities by using the
Abbot-Deser-Tekin (ADT) method~\cite{Abbott:1981ff,Deser:2002jk}.
The ADM method is suitable for computing conserved quantities of a
black hole in asymptotically flat spacetimes, while the ADT method
is useful to compute conserved quantities of a black hole in
asymptotically AdS spacetimes found from fourth-order
gravity~\cite{Kim:2013zha}. After computing all ADT thermodynamic
quantities depending on a mass parameter $m^2_d(=1/\beta)$,  one is
ready to study thermodynamics and phase transition between TAdS and
AdS black hole in fourth-order gravity. For $m^2_d>m^2_{\rm c}$ with
critical mass parameter $m_c$ giving ${\cal M}_d^2(m^2_c)=0$, all
thermodynamic properties are dominantly determined by Einstein
gravity, while for $m^2_d<m^2_{\rm c}$, all thermodynamic properties
are dominantly by Wely-squared term. The former is completely
understood, but the latter becomes a new area of black hole
thermodynamics appeared when one studies the black hole  by using
the ADT thermodynamic quantities.

On the other hand, there was a connection between thermodynamic
instability and classical [Gregory-Laflamme
(GL)~\cite{Gregory:1993vy}] instability for the black
strings/branes. This  Gubser-Mitra proposal~\cite{Gubser:2000ec} was
referred to as the the correlated stability conjecture
(CSC)~\cite{Harmark:2007md} which states that the classical
instability of a black string/brane with translational symmetry and
infinite extent sets in precisely when the corresponding
thermodynamic system becomes (locally) thermodynamically unstable
(Hessian matrix\footnote{There are two representations when defining
the Hessian matrix: $H^{S}_{M}$ and $H^{M}_{S}$. The matrix
$H^{S}_{M}$ ($H^{M}_{S}$) can be expressed in terms of the
second-order derivatives of the entropy (mass) with respect to the
mass (entropy) and the conserved charges.  Here, Hessian matrix $<0$
denotes  a negative eigenvalue of the matrix $H^{M}_{S}$.} $<0$ or
heat capacity $<0$). Here the additional assumption of translational
symmetry and infinite extent has been added to ensure that finite
size effects do not spoil the thermodynamic nature of the argument
and to exclude a well-known case of  the Schwarzschild black hole
which is classically stable, but thermodynamically unstable because
of its negative heat capacity.

Interestingly, it is very important to mention that the stability of
the Schwarzschild black hole in four-dimensional massive gravity  is
determined by using the GL instability of a five-dimensional black
string.   Although the Schwarzschild black hole stability has been
performed in Einstein gravity forty years ago
~\cite{Regge:1957td,Zerilli:1970se,Vishveshwara:1970cc}, the
stability analysis of the Schwarzschild black hole in massive
gravity theory were very recently announced. The massless spin-2
graviton has 2 degrees of freedom (DOF) in Einstein gravity, while
the massive graviton has 5 DOF in massive gravity theory. Even a
massive spin-2  graviton has 5 DOF, one  has a single physical DOF
when one considers the $s(l=0)$-mode of massive graviton. Also, it
was proved that the $s$-wave perturbation gives unstable modes only
in the higher dimensional black string
perturbation~\cite{Kudoh:2006bp}. It turned out that the small
Schwarzschild black holes in the dRGT massive
gravity~\cite{Babichev:2013una,Brito:2013wya} and fourth-order
gravity~\cite{Myung:2013doa,Myung:2013cna} are unstable against the
metric and Ricci tensor perturbations, respectively. This implies
that the massiveness of $m^2\not=0$ gives rise to unstable modes
propagating around the Schwarzschild black hole.  If one may  find
thermodynamic instability from the ADT thermodynamic quantities of
AdS black hole in fourth-order gravity, then it could be  compared
with the GL-instability found from the linearized Einstein
equation~\cite{Myung:2013bow}.  If one finds  a connection between
them, it might imply that the Gubser-Mitra conjecture  holds  even
for a compact object of the SAdS black hole found in fourth-order
gravity. This is our main motivation of why we study fourth-order
gravity here.

In this work, we investigate  thermodynamic and classical
instability of AdS black holes in fourth-order gravity. These
include the BTZ black hole in new massive gravity, Schwarzschild-AdS
black hole and higher-dimensional AdS black holes in fourth-order
gravity.  All thermodynamic quantities are computed using the ADT
method. Here we use the ADT conserved quantities, since they respect
the first-law of thermodynamics and the ADT mass and entropy are
reliable  to use  a thermodynamic study of the AdS black holes in
fourth-order gravity.  Finally, we establish  a connection between
the thermodynamic instability of AdS black holes and the GL
instability of AdS black holes in the linearized fourth-order
gravity.

\section{BTZ black hole in new massive gravity}

As a prototype, we consider the BTZ black hole in new massive
gravity (NMG) which is known to be  a three-dimensional version of
fourth-order gravity.  The NMG action~\cite{Bergshoeff:2009hq}
composed of the Einstein-Hilbert action with a cosmological constant
$\lambda$ and fourth-order curvature terms is given by
\begin{eqnarray}
\label{NMGAct}
 S_{\rm NMG} &=&S_{\rm EH}+S_{\rm FOT },  \\
 \label{NEH} S_{\rm EH}&=& \frac{1}{16\pi G_3} \int d^3x \sqrt{-g}~
  (R-2\lambda), \\
\label{NFO} S_{\rm FOT }&=&-\frac{1}{16\pi G_3m^2} \int d^3x
            \sqrt{-g}~\left(R_{\mu\nu}R^{\mu\nu}-\frac{3}{8}R^2\right),
\end{eqnarray}
where $G_3$ is a three-dimensional Newton constant and $m^2$ a
positive mass parameter with mass dimension 2 [$m^2 \in
(0,\infty)$]. In the limit of $m^2 \to \infty$, $S_{\rm NMG}$
recovers  the Einstein gravity  $S_{\rm EH}$, while it reduces to
purely fourth-order  term  $S_{\rm FOT }$ in the limit of $m^2 \to
0$. The field equation is given by \be \label{eqn}
R_{\mu\nu}-\frac{1}{2}g_{\mu\nu}R+\lambda
g_{\mu\nu}-\frac{1}{2m^2}K_{\mu\nu}=0,\ee where
\begin{eqnarray}
  K_{\mu\nu}&=&2\square R_{\mu\nu}-\frac{1}{2}\nabla_\mu \nabla_\nu R-\frac{1}{2}\square{}Rg_{\mu\nu}\nonumber\\
        &+&4R_{\mu\rho\nu\sigma}R^{\rho\sigma} -\frac{3}{2}RR_{\mu\nu}-R_{\rho\sigma}R^{\rho\sigma}g_{\mu\nu}
         +\frac{3}{8}R^2g_{\mu\nu}.
\end{eqnarray}
The non-rotating BTZ  black hole solution to Eq.(\ref{eqn}) is given
by~\cite{BTZ-1,BTZ-2}
 \be
\label{2dmetric}
  ds^2_{\rm BTZ}=\bar{g}_{\mu\nu}dx^\mu dx^\nu=-f(r)dt^2
   +\frac{dr^2}{f(r)}+r^2d\phi^2,~~f(r)=-M+\frac{r^2}{\ell^2}
\end{equation}
under the condition of $1/\ell^2+\lambda+1/(4m^2\ell^4)=0$. Here $M$
is an integration constant related to the the ADM mass of black
hole. The horizon radius $r_+$ is determined by the condition of
$f(r)=0$ and $\ell$ denotes the curvature radius of AdS$_3$
spacetimes.

Its Hawking  temperature is found to be  \be T_{\rm H}=
\frac{f'(r_+)}{4\pi}=\frac{r_+}{2\pi \ell^2}. \ee Using the ADT
method, one can derive  all thermodynamic quantities of  its
mass~\cite{Clement:2009gq}, heat capacity ($C=\frac{dM_{\rm
ADT}}{dT_{\rm H}}$), entropy~\cite{Kim:2013qra}, and on-shell
(Helmholtz) free energy
\begin{eqnarray}  && M_{\rm
ADT}(m^2,r_+)=\Big(1-\frac{1}{2m^2\ell^2}\Big)M(r_+),~~C_{\rm
ADT}(m^2,r_+)=\Big(1-\frac{1}{2m^2\ell^2}\Big)C(r_+), \nonumber \\
&&\label{lbh} S_{\rm
ADT}(m^2,r_+)=\Big(1-\frac{1}{2m^2\ell^2}\Big)S_{\rm
BH}(r_+),~~F^{on}_{\rm
ADT}(m^2,r_+)=\Big(1-\frac{1}{2m^2\ell^2}\Big)F^{on}(r_+),\end{eqnarray}
whose thermodynamic quantities in Einstein gravity have already
given by~\cite{Myung:2005ee,Myung:2006sq,Eune:2013qs} \be
\label{btz} M(r_+)=\frac{r_+^2}{8G_3\ell^2},~C(r_+)=\frac{\pi
r_+}{2G_3},~S_{BH}(r_+)=
 \frac{\pi r_+}{2G_3},~F^{on}(r_+)= M-T_{\rm H} S_{BH}=-\frac{r_+^2}{8G_3\ell^2}.\ee
These all are positive regardless of the horizon size $r_+$ except
that the free energy is always negative. This means that the BTZ
black hole is thermodynamically stable in Einstein gravity.  Here we
check that the first-law of thermodynamics is satisfied as
\be\label{first-law}dM_{\rm ADT}=T_{\rm H} dS_{\rm ADT} \ee as the
first-law is satisfied in Einstein gravity
 \be
dM=T_{\rm H} dS_{\rm BH} \ee where `$d$' denotes the differentiation
with respect to the horizon size $r_+$ only. In this work, we treat
$m^2$ differently from the black hole charge $Q$ and angular
momentum $J$ to achieve the first-law (\ref{first-law}).  Here we
observe that in the limit of $m^2 \to \infty$ one recovers
thermodynamics of the BTZ black hole in Einstein gravity, while in
the limit of $m^2 \to 0$ we recover the black hole thermodynamics in
purely fourth-order gravity which is similar to recovering the
conformal Chern-Simons gravity  from the topologically massive
gravity (TMG)~\cite{Bagchi:2013lma}.

On the other hand, the linearized equation to (\ref{eqn}) upon
choosing the transverse-traceless (TT)  gauge of $\bar{\nabla}^\mu
h_{\mu\nu}=0$ and $h^\mu~_\mu=0$ leads to the fourth-order equation
for the metric perturbation
$h_{\mu\nu}$~\cite{Myung:2011bn,Moon:2013jna}
\begin{equation}
\Big(\bar{\nabla}^2-2\Lambda\Big) \Big[\bar{\nabla}^2-2\Lambda
-{\cal M}^2(m^2)\Big] h_{\mu\nu} =0,~~\Lambda=-\frac{1}{\ell^2}
\label{linh}
\end{equation}
which might imply  the two second-order linearized equations
\begin{eqnarray}\label{nmgmeq1}
&&\Big(\bar{\nabla}^2-2\Lambda\Big)h_{\mu\nu}=0,
\\ \label{nmgmeq2}
&&\Big[\bar{\nabla}^2-2\Lambda -{\cal M}^2(m^2)\Big]h_{\mu\nu}=0.
\end{eqnarray}
Here the mass squared of a massive spin-2 graviton is given by \be
{\cal M}^2(m^2)=m^2-\frac{1}{2\ell^2}. \ee Eq. (\ref{nmgmeq2})
describes a massive graviton with 2 DOF propagating around the BTZ
black hole under the TT gauge.

Expressing all thermodynamic quantities in (\ref{lbh}) in terms of
the mass squared leads to
\begin{eqnarray} \label{lbh2} M_{\rm
ADT}=\frac{{\cal M}^2}{m^2}M,~~C_{\rm ADT}=\frac{{\cal
M}^2}{m^2}C,~~ S_{\rm ADT}=\frac{{\cal M}^2}{m^2}S_{\rm
BH},~~F^{on}_{\rm ADT}=\frac{{\cal M}^2}{m^2}F^{on}\end{eqnarray}
which shows clearly that all thermodynamical quantities depend on
the sign of ${\cal M}^2$. The local thermodynamic stability is
determined by the positive heat capacity ($C_{\rm ADT}>0$) and the
global stability is determined by the negative free energy ($F_{\rm
ADT}<0$). Hence, it implies that the thermodynamic stability is
determined by the sign of the heat capacity, while the phase
transition is mainly determined by the sign of the free energy.

 For ${\cal M}^2>0(m^2>1/2\ell^2)$, all thermodynamic
quantities have the same property as those for Einstein gravity
(\ref{NEH}), whereas for ${\cal M}^2<0(m^2<1/2\ell^2)$, all
thermodynamic quantities have the same property as those for
fourth-order term (\ref{NFO}).   We observe from Fig. 1 that for
${\cal M}^2>0$, the BTZ black hole is thermodynamically stable
regardless of the horizon size $r_+$ because of  $C_{\rm ADT}>0$.

 On the other hand, the classical (in)
stability condition of the BTZ black hole was recently determined by
the condition of ${\cal M}^2>0(<0)$ regardless of the horizon size
$r_+$~\cite{Moon:2013lea}.  The case of ${\cal M}^2=0$ corresponds
to the critical gravity where all thermodynamical quantities are
zero and logarithmic modes appear. For ${\cal M}^2<0$, the BTZ black
hole is thermodynamically unstable because of $C_{\rm ADT}<0$  as
well as it is classically unstable against the metric perturbations.
{\it Hence, it shows a clear connection between thermodynamic and
classical instability for the BTZ black hole regardless of the
horizon size in new massive gravity.}
\begin{figure}[t!]
   \centering
   \includegraphics{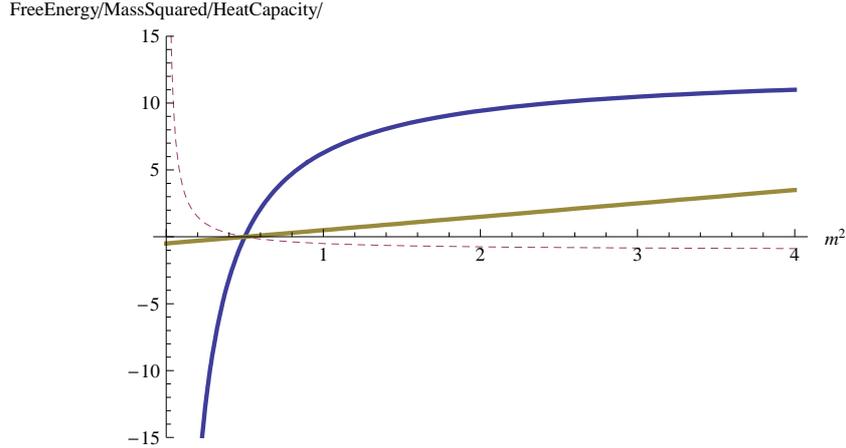}
\caption{ Heat capacity (solid curve) $C_{\rm ADT}(m^2,r_+=1)$, mass
squared (line) ${\cal M}^2(m^2)$, and free energy (dotted curve)
$F^{on}_{\rm ADT}(m^2,r_+=1)$ with $G_3=1/8$ and $\ell=1$. For
$m^2<0.5$, one has $C_{\rm ADT}<0(F^{on}_{\rm ADT}>0)$ and ${\cal
M}^2<0$, while for $m^2>0.5$, one has $C_{\rm ADT}>0(F^{on}_{\rm
ADT}<0)$ and ${\cal M}^2>0$. At the critical point of $m^2=0.5$, we
have $F^{on}_{\rm ADT}=0$ and ${\cal M}^2=0$. In the limit of
$m^2\to 0/\infty$, fourth-order term/Einstein gravity are recovered
for free energy and heat capacity.}
\end{figure}

Finally, let us turn to the issue related to a phase transition from
the thermal AdS$_3$ (TAdS) to BTZ black hole.  For this purpose, we
first consider thermodynamic quantities for the
TAdS~\cite{Myung:2006sq} \be \label{tads} M_{\rm
TAdS}=-\frac{1}{8G_3},~F_{\rm TAdS}=-\frac{1}{8G_3}.\ee It turns out
that for ${\cal M}^2<0(m^2<1/2\ell^2)$ [fourth-order term
(\ref{NFO}) contributes dominantly to black hole thermodynamics],
the TAdS is always favored than the BTZ black hole because of
$F_{\rm TAdS}<F_{\rm ADT}^{on}$, where $F_{\rm ADT}^{on}$ is given
by (\ref{lbh2}). In this case, we might not define a possible phase
transition because the ground state is the TAdS. Alternatively, this
implies that a gab between $F^{on}_{\rm ADT}$ and $F_{\rm TAdS}$
might not allow a continuous phase transition\footnote{Concerning
this issue, a phase transition between hot flat space and flat space
cosmological spacetimes was recently studied in TMG  by using
on-shell free energies~\cite{Bagchi:2013lma}.}. It is noted that a
phase transition from the TAdS to the BTZ black hole is possible to
occur in new massive gravity for ${\cal M}^2>0(m^2=1>1/2\ell^2)$
(see arXiv:1311.6985v1 for details), in which the Einstein gravity
(\ref{NEH}) contributes dominantly to black hole thermodynamics. The
corresponding phase transition in this case is similar to that
obtained in the literature \cite{Myung:2012xc} for $z=1$.

\section{Thermodynamics of AdS black holes in fourth-order gravity}

Let us start with the $d(\ge4)$-dimensional fourth-order gravity
action~\cite{Liu:2011kf}
\begin{eqnarray}
S_{\rm dFO}=\frac{1}{16\pi G_d}\int d^d x\sqrt{-g} \Big[R
-(d-2)\Lambda_0+\alpha R_{\mu\nu}R^{\mu\nu}+\beta R^2\Big]
\label{Action}
\end{eqnarray} with two parameters $\alpha$ and $\beta$.
Here we  do not include the  Gauss-Bonnet term~\cite{Deser:2011xc}
because (\ref{Action}) admits solutions of the higher-dimensional
Einstein gravity including the higher dimensional AdS black holes.
From (\ref{Action}), the Einstein equation is derived to be
\begin{equation} \label{equa1}
G_{\mu\nu}+E_{\mu\nu}=0,
\end{equation}
where the Einstein tensor  is given by \begin{equation}
G_{\mu\nu}=R_{\mu\nu}-\frac{1}{2} Rg_{\mu\nu}+\frac{d-2}{2}\Lambda_0
g_{\mu\nu}
\end{equation}
and $E_{\mu\nu}$  takes the form
\begin{eqnarray} \label{equa2}
E_{\mu\nu}&=& 2\alpha
\Big(R_{\mu\rho\nu\sigma}R^{\rho\sigma}-\frac{1}{4}
R^{\rho\sigma}R_{\rho\sigma}g_{\mu\nu}\Big)+2\beta
R\Big(R_{\mu\nu}-\frac{1}{4} Rg_{\mu\nu}\Big) \nonumber \\
&+&
\alpha\Big(\nabla^2R_{\mu\nu}+\frac{1}{2}\nabla^2Rg_{\mu\nu}-\nabla_\mu\nabla_\nu
R\Big) +2\beta\Big(g_{\mu\nu} \nabla^2R-\nabla_\mu \nabla_\nu
R\Big).
\end{eqnarray}
For the Einstein space of $R_{\mu\nu}=\Lambda g_{\mu\nu}$ and
$R=d\Lambda$ together with
$\Lambda_0=\Lambda+(d-4)(\alpha+d\beta)\Lambda^2/(d-2)$,
Eq.(\ref{equa1}) allows a $d$-dimensional AdS black hole solution
\begin{equation} \label{sch} ds^2_{\rm ST}=\bar{g}_{\mu\nu}dx^\mu
dx^\nu=-V(r)dt^2+\frac{dr^2}{V(r)}+r^2d\Omega^2_{d-2}
\end{equation}
with the metric function \begin{equation} \label{num}
V(r)=1-\Big(\frac{r_0}{r}\Big)^{d-3}-\frac{\Lambda}{d-1}r^2,~~\Lambda=-\frac{d-1}{\ell^2}.
\end{equation} The horizon is located at $r=r_+$ ($V(r_+)=0$) which means that $r_0$ differs from $r_+$.
Hereafter we denote the background quantities with
the ``overbar''. In this case, the black hole background spacetimes
is given by
\begin{equation}
\bar{R}_{\mu\nu}=\Lambda\bar{g}_{\mu\nu},~~\bar{R}=4\Lambda.
\end{equation}

The Hawking  temperature is derived as \be T^d_{\rm H}=
\frac{V'(r_+)}{4\pi}=\frac{1}{4\pi r_+}\Big[(d-3)
+\frac{d-1}{\ell^2}r_+^2\Big]. \ee  Using the ADT
method~\cite{Abbott:1981ff,Deser:2002jk}, all thermodynamic
quantities of  its mass~\cite{Liu:2011kf}, heat capacity,
entropy~\cite{Liu:2011kf}, and on-shell free energy are given by
\begin{eqnarray}  && M_{\rm
dADT}=\frac{4(d-1)}{dm_d^2}{\cal M}_d^2 M_d(r_+),~~C_{\rm
dADT}(m^2,r_+)=\frac{4(d-1)}{dm_d^2}{\cal M}_d^2C_d(r_+), \nonumber \\
&&\label{dlbh} S_{\rm dADT}=\frac{4(d-1)}{dm_d^2}{\cal M}_d^2S_{\rm
BH}(r_+),~~F^{on}_{\rm dADT}=\frac{4(d-1)}{dm_d^2}{\cal M}_d^2
F^{on}_d(r_+),\end{eqnarray} where \be m^2_d=\frac{1}{\beta},~~
\alpha=-\frac{4(d-1)}{d} \beta,~~{\cal
M}^2_d(m^2_d)=\frac{d}{4}\Big[\frac{m_d^2}{d-1}-\frac{2(d-2)^2}{d\ell^2}\Big].
\ee At this stage, we note that even though all thermodynamic
quantities are obtained  for arbitrary $\alpha $ and $\beta$, we
require a condition of $\alpha=-4(d-1)\beta/d$ because the classical
stability could be achieved  only under this condition. This means
that we have a single mass parameter $m^2_d=1/\beta$ by eliminating
a massive spin-0 graviton in fourth-order gravity, reducing to the
Einstein-Weyl gravity.
 All
thermodynamic quantities in $d(\ge 4)$-dimensional Einstein gravity
were known to be ~\cite{Myungtads}
\begin{eqnarray} \label{hbtz1}
M_d(r_+)&=&\frac{\Omega_{d-2}(d-2)}{16\pi
G_d}r_+^{d-3}\Big[1+\frac{r_+^2}{\ell^2}\Big], \\
C_d(r_+)&=&\frac{dM_d}{dT^d_{\rm
H}}=\frac{\Omega_{d-2}(d-2)r_+^{d-2}}{4G_d}
\Big[\frac{(d-1)r_+^2+(d-3)\ell^2}{(d-1)r_+^2-(d-3)\ell^2}\Big],\label{hbtz2}
\\~S_{BH}(r_+)&=&
 \frac{\Omega_{d-2}}{4G_d}r^{d-2}_+,\label{hbtz3}\\
 F^{on}_d(r_+)&=& M_d-T_{\rm H}
 S_{BH}=\frac{\Omega_{d-2}}{16\pi G_d}r_+^{d-3}\Big[1-\frac{r_+^2}{\ell^2}\Big]
  \label{hbtz4}\end{eqnarray}
with the area of $S^{d-2}$ \be
\Omega_{d-2}=\frac{2\pi^{\frac{d-1}{2}}}{\Gamma(\frac{d-1}{2})}. \ee
We observe from (\ref{dlbh}) that for ${\cal M}^2_d>0$,
thermodynamic stability of
 AdS black hole is determined by  the
higher-dimensional Einstein gravity. On the other hand, for  ${\cal
M}^2_d<0$, thermodynamic stability of black hole is determined by
Weyl-squared term (conformal gravity).

We check that the first-law of thermodynamics is satisfied as
\be\label{dfirst-law}dM_{\rm dADT}=T^d_{\rm H} dS_{\rm dADT} \ee as
the first-law is satisfied in $d$-dimensional  Einstein gravity
 \be
dM_d=T^d_{\rm H} dS_{\rm BH}, \ee where `$d$' denotes the
differentiation with respect to the horizon size $r_+$ only. In this
work, we treat $m^2_d$ differently from the black hole charge $Q$
and angular momentum $J$  to obtain the first-law
(\ref{dfirst-law}). Here we observe that in the limit of $m^2_d \to
\infty$ we recovers thermodynamics of the AdS black hole in
$d$-dimensional  Einstein gravity, while in the limit of $m^2_d \to
0$ we recover that in Weyl-squared term (conformal gravity).

\section{SAdS black hole in Einstein-Weyl gravity}

\subsection{Thermodynamic instability for small black holes}
\begin{figure}[t!]
   \centering
   \includegraphics{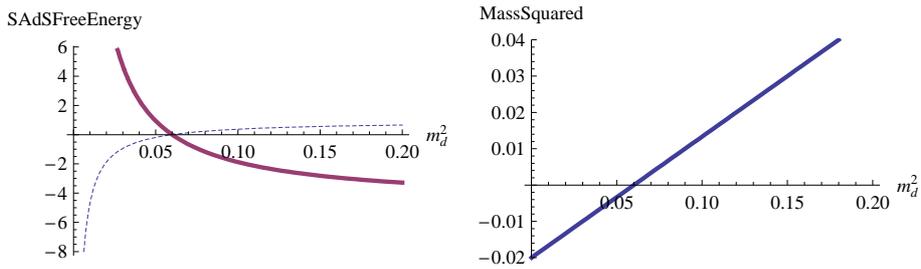}
\caption{Left: Free energy (solid) $F^{on}_{\rm dADT}(m^2_d,r_+=15)$
for large black hole and  free energy (dotted) $F^{on}_{\rm
dADT}(m^2_d,r_+=5)$ for small black hole with $G_4=1/2$ and
$\ell=10$. Right: mass squared ${\cal M}^2_d(m^2_d)$.   At the
critical point of $m^2_d=0.06$, we have $F^{on}_{\rm dADT}=0$ and
${\cal M}^2_d=0$. In the limit of $m^2_d\to 0/\infty$, Weyl-squared
term (conformal gravity)/Einstein gravity are recovered for free
energy.}
\end{figure}

For a definite description of black hole thermodynamics, we choose
$d=4$ which provides a SAdS black hole. Its thermodynamic quantities
of mass~\cite{Lu:2011zk}, heat capacity, entropy~\cite{Lu:2011zk},
and on-shell free energy are given by
\begin{eqnarray}
M_{\rm dADT}(m^2_d,r_+)&=&\Big(1-\frac{6}{m^2_d\ell^2}\Big)M_{\rm SAdS},\\
C_{\rm dADT}(m^2_d,r_+)&=& \Big(1-\frac{6}{m^2_d\ell^2}\Big)C_{\rm SAdS}, \\
S_{\rm dADT}(m^2_d,r_+)&=& \Big(1-\frac{6}{m^2_d\ell^2}\Big)S_{\rm BH}, \\
F^{on}_{\rm dADT}(m^2_d,r_+)&=&
\Big(1-\frac{6}{m^2_d\ell^2}\Big)F^{on}_{\rm SAdS},
\end{eqnarray}
where all thermodynamic quantities of SAdS black hole are shown in
the Eqs. (\ref{hbtz1})-(\ref{hbtz4}) for $d=4$.  The mass squared
takes the form~\cite{Lu:2011zk} \be {\cal
M}^2_d=\frac{m^2_d}{3}-\frac{2}{\ell^2} \ee which is
negative/postive for $m^2_d\lessgtr 6/\ell^2$.

First of all, we depict the free energy as a function of $m^2_d$ in
 Fig. 2. Since the sign of free energy depends on the horizon size
$r_+$ critically, we plot the two free energies as  function of
$m^2_d$ for large ($r_+=15>\ell=10)$ and small black hole
($r_+=5<\ell$), respectively.
 For $m^2_d<0.06$, one has $F^{on}_{\rm
dADT}(m^2_d,r_+=15)>0(F^{on}_{\rm dADT}(m^2_d,r_+=5)<0)$ and ${\cal
M}^2_d<0$, while for $m^2_d>0.06$, one has $F^{on}_{\rm
dADT}(m^2_d,r_+=15)<0(F^{on}_{\rm dADT}(m^2_d,r_+=5)>0)$ and ${\cal
M}^2_d>0$.

We consider first the case of ${\cal M}^2_d>0(m^2_d>0.06)$ which is
dominantly described by the Einstein gravity.  Since the heat
capacity of $C_{\rm SAdS}$ blows up at
$r_+=r_*=\ell/\sqrt{3}=0.577\ell$, we divide the black hole into the
small black hole with $r_+<r_*$ and the large  black hole with
$r_+>r_*.$  As is shown in
 the solid (familiar) curves in Fig. 3,  we have the small black hole with $r_+<r_*$ which is thermodynamically
unstable because $C_{\rm dADT} <0$, while the large black hole with
$r_+>r_*$ is thermodynamically  stable because $C_{\rm dADT}
>0$.
\begin{figure}[t!]
   \centering
   \includegraphics{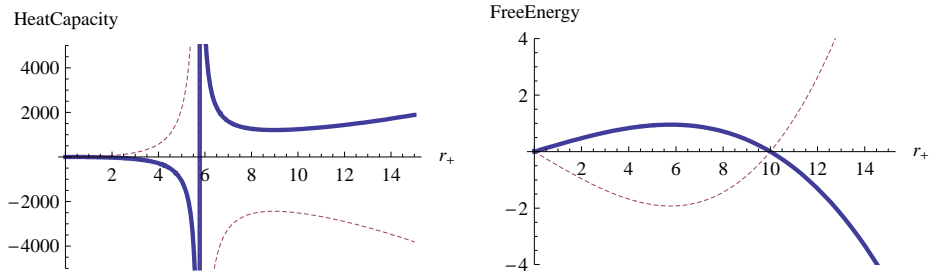}
\caption{Left: Familiar  heat capacity (solid) $C_{\rm
dADT}(m^2_d=6,r_+)$ for ${\cal M}^2_d=1.98>0$ and unfamiliar heat
capacity (dotted) $C_{\rm dADT}(m^2_d=0.02,r_+)$ for ${\cal
M}^2_d=-0.013<0$ with $G_4=1/2$ and $\ell=10$. Right: Familiar free
energy (solid) $F^{on}_{\rm dADT}(m^2_d=6,r_+)$ for ${\cal
M}^2_d=1.98$ and unfamiliar free energy (dotted) $F^{on}_{\rm
dADT}(m^2_d=0.02,r_+)$ for ${\cal M}^2_d=-0.013$. }
\end{figure}
Especially for ${\cal M}^2_d=1.98>0(m^2_d=6)$, one has $M_{\rm
dADT}=0.99M_{\rm SAdS},~C_{\rm dADT}=0.99C_{\rm SAdS},~S_{\rm
dADT}=0.99S_{\rm BH}$, and $F_{\rm dADT}=0.99F_{\rm SAdS}$.  Hence
we could describe the Hawking-Page phase transition well as for the
SAdS black hole in Einstein gravity~\cite{HP}. We wish to comment
that   the free energy has the maximum value at $r_+=r_*$. Since the
free energy becomes negative (positive) for $r_+<\ell(r_+>\ell)$, we
did not choose $r_+=\ell$ as a boundary point to divide the black
hole into small and large black holes.

On the other hand, for  ${\cal M}^2_d=-0.013<0(m^2_d=0.02<6/\ell^2)$
which is dominantly described by conformal gravity~\cite{Lu:2012xu},
the small black hole ($r_+<r_*$) is thermodynamically stable because
$C_{\rm dADT}
>0$, while the large black hole
($r_+>r_*$) is thermodynamically  unstable because $C_{\rm dADT}
<0$.  See the dotted (unfamiliar) curves in Fig. 3 for observation.
It seems that there is no known phase transition from thermal AdS to
the SAdS black hole in conformal  gravity.

\subsection{GL instability for small black holes}

We briefly review the Gregory-Laflamme $s$-mode instability for a
massive spin-2 graviton  with mass ${\cal M}_d \ge 0$ propagating on
the SAdS black hole spacetimes in Einstein-Weyl gravity. Choosing
the TT gauge, its linearized equation to (\ref{equa1}) takes the
form
\begin{equation} \label{lineqq}
\bar{\nabla}^2
h_{\mu\nu}+2\bar{R}_{\alpha\mu\beta\nu}h^{\alpha\beta}-{\cal
M}^2_dh_{\mu\nu}=0.
\end{equation}
which describes  5 DOF of a massive spin-2 graviton propagating on
the SAdS black hole spacetimes. We note that  choosing the condition
of $\alpha=-3\beta$ eliminates a massive spin-0 graviton with 1 DOF.

Before we proceed, we wish to  mention that the stability of the
Schwarzschild black hole in four-dimensional massive gravity is
determined by using the Gregory-Laflamme instability of a
five-dimensional black string. It turned out that the small
Schwarzschild black holes in the dRGT massive
gravity~\cite{Babichev:2013una,Brito:2013wya} and fourth-order
gravity~\cite{Myung:2013doa} are  unstable against the metric and
Ricci tensor perturbations because the inequality is satisfied as
\begin{equation}
{\cal M}_d \le \frac{{\cal O}(1)}{r_0},~~r_0=2M_S.
\end{equation}
For the massless case of  ${\cal M}_d=0$, Eq. (\ref{lineqq}) leads
to the linearized equation around the Schwarzschild black hole  with
the TT gauge which is known to be stable in the Einstein gravity.

Choosing the $s$-mode ansatz whose form is given by
$H_{tt},~H_{tr},~H_{rr},$ and $K$ as
\begin{eqnarray}
h^s_{\mu\nu}=e^{\Omega t} \left(
\begin{array}{cccc}
H_{tt}(r) & H_{tr}(r) & 0 & 0 \cr H_{tr}(r) & H_{rr}(r) & 0 & 0 \cr
0 & 0 &  K(r) & 0 \cr 0 & 0 & 0 & \sin^2\theta K(r)
\end{array}
\right), \label{evenp}
\end{eqnarray}
a relevant equation for $ H_{tr}$ takes the same form (see Appendix for explicit forms of $A,~B,~C$)
\begin{equation} \label{secondG-eq} A(r;r_0,\ell,\Omega^2,{\cal M}_d^2)
\frac{d^2}{dr^2}H_{tr} +B\frac{d}{dr}H_{tr}+CH_{tr}=0,
\end{equation}
which shows the same unstable modes for
\begin{equation} \label{unst-con}
0<{\cal M}_d<\frac{{\cal O}(1)}{r_0} \end{equation} with the mass
\begin{equation} {\cal M}_d=\sqrt{\frac{m^2_d}{3}-\frac{2}{l^2}}.
\end{equation}
The condition of  (\ref{unst-con})  could be  read off from Fig. 4
when one notes the difference between $r_+$ and $r_0$: $r_+=1,2,4$
and $r_0=1.01,2.08,4.64$.
\begin{figure}[t!]
   \centering
   \includegraphics{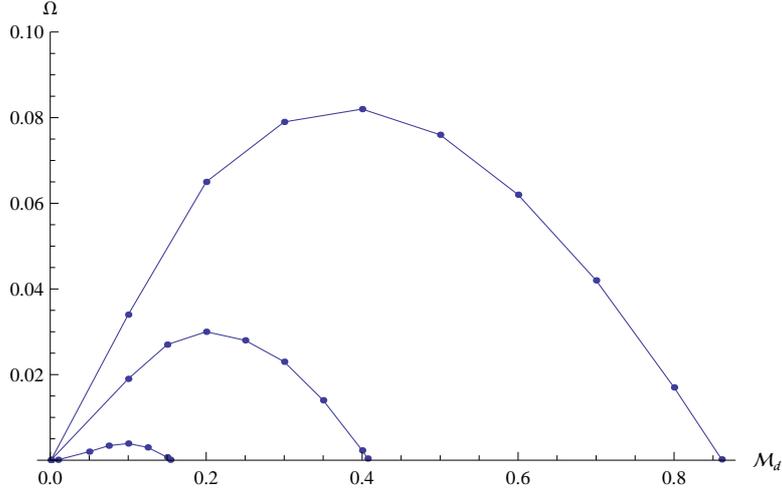}
\caption{Plots of unstable modes on three curves with $r_+=1,2,4$
and $\ell=10$. These belong to small black holes because of
$r_+<r_*=5.77$. The $y(x)$-axis denote $\Omega({\cal M}_d)$. Also we
check that for $r_+=6>r_*=5.77$, the maximum value of $\Omega$ is
less than $10^{-4}$, which implies that there is no instability for
large black hole. In this figure, the smallest curve represents
$r_+=4$, the medium denotes $r_+=2$, and the largest one shows
$r_+=1$.} \label{fig.6}
\end{figure}
On the other hand, the stable condition  of the SAdS black hole in
Einstein-Weyl gravity is given by
\begin{equation} \label{st-con} {\cal M}_d>\frac{{\cal O}(1)}{r_0}.
\end{equation}

At this stage,  we would like to mention the classical stability of
${\cal M}_d=0$ case. In this case, its linearized equation reduces
to
\begin{equation}
\bar{\nabla}^2
h_{\mu\nu}+2\bar{R}_{\alpha\mu\beta\nu}h^{\alpha\beta}=0,
\end{equation}
which is exactly the linearized equation around the SAdS black hole
in Einstein gravity. From the observation\footnote{In the next
section 5.2, we introduce the corresponding numerical analysis to
observe this GL instability.} of Fig. 4, the GL instability
disappears at ${\cal M}_d=0$, which may imply that the SAdS black
hole is stable against the $s$-mode metric perturbation. The SAdS
black hole was known to be stable against the metric perturbation
even though a negative potential appeared near the event horizon in
odd-parity sector~\cite{Cardoso:2001bb}. Later on, one could achieve
the positivity of gravitational potentials  by using the
$S$-deformed technique~\cite{Ishibashi:2003ap},  proving the
stability of SAdS black hole exactly~\cite{Moon:2011sz}. This
implies that there is no connection between classical stability and
thermodynamic instability ($C_{\rm SAdS}<0$) for small SAdS black
hole.  This situation is similar to the Schwarzschild black hole
which shows a violation of the CSC between thermodynamic instability
($C_{\rm S}<0$) and the classical
stability~\cite{Regge:1957td,Zerilli:1970se,Vishveshwara:1970cc}.
Therefore, we could not apply the Gubser-Mitra conjecture to the
SAdS black hole in Einstein gravity.

Let us see how things are improved in Einstein-Weyl gravity. We note
that at the critical point of ${\cal M}^2_d=0$, all thermodynamic
quantities vanish exactly. For ${\cal M}^2_d
>0$ and small black hole with $r_+ <r_*$, the heat capacity  takes
the form
\begin{equation} \label{ther-con}C_{\rm dADT}=\frac{3{\cal
M}^2_d}{m^2_d}C_{\rm SAdS}<0,
\end{equation}
which shows thermodynamic instability like that of a small SAdS
black hole in Einstein gravity. From the condition of
(\ref{unst-con}), however, we find that a small black hole is
unstable against the $s$-mode massive graviton perturbation. This
implies that the CSC holds  for the SAdS black hole in Einstein-Weyl
gravity.

 Also the stability
condition of (\ref{st-con}) is consistent with thermodynamic
stability condition for large black hole with $r_+>r_*$ in Einstein
gravity
\begin{equation}
\label{ther-con}C_{\rm dADT}=\frac{3{\cal M}^2_d}{m^2_d}C_{\rm
SAdS}>0.
\end{equation}
 As was previously
emphasized, there is no connection between thermodynamic instability
and classical stability for small SAdS black hole in Einstein
gravity. However, the GL instability condition picks up the small
SAdS black hole which is thermodynamically unstable in Einstein-Weyl
gravity. Hence, we conclude that there is  a connection between the
GL instability and thermodynamic instability for small black hole in
fourth-order (Einstein-Weyl) gravity.

\section{Higher-dimensional AdS black holes
in fourth-order gravity}
\subsection{Thermodynamic instability for small black holes}
In this section, we comment briefly on the thermodynamic
(in)stability for higher-dimensional AdS black hole. To this end, we
first recall the thermodynamic quantities (\ref{dlbh}), obtained in
the $d$-dimensional fourth order gravity. Among them, taking into
account the heat capacity together with (\ref{hbtz2}), the small and
large black holes can be divided  by choosing  the blow-up heat
capacity at
\begin{equation}
r_+=r^{(d)}_*=\sqrt{\frac{d-3}{d-1}}\ell. \end{equation}

For ${\cal M}^2_d>0$ [$m^2_d>2(d-1)(d-2)^2/d\ell^2]$, we have the
small black hole for $r_+<r^{(d)}_*$ which is thermodynamically
unstable because $C_{\rm dADT}<0$ in (\ref{dlbh}), while we have the
large black hole for $r_+>r^{(d)}_*$ which  is thermodynamically
stable because $C_{\rm dADT}>0$. This is dominantly described by the
higher-dimensional Einstein gravity. We would like to  mention that
for ${\cal M}^2_d>0$ we will  establish the connection between the
GL instability and thermodynamic instability of the small black
hole.

On the other hand, for ${\cal M}^2_d<0$
[$m^2_d<2(d-1)(d-2)^2/d\ell^2]$ which is dominantly described by
Weyl-squared term, the small black hole is thermodynamically stable
because $C_{\rm dADT}>0$, whereas the large black hole is
thermodynamically unstable because of  $C_{\rm dADT}<0$. This case
requires  a newly black hole thermodynamics.

\subsection{GL instability}
In order to investigate the classical instability for
higher-dimensional AdS black hole, we first consider two coupled
first order differential equations\footnote{We note that these first
order differential and constraint equations can be obtained from
using the perturbation equation(\ref{lineqq}) and TT gauge
condition. Finally we have checked, after some manipulations, that
these equations are consistent with the second order equation
(\ref{secondG-eq}) and for $V=1-(r_0/r)^{d-3}$ [in the  $\ell^2\to
\infty$-limit], they reduce to those found in the original
literature~\cite{Gregory:1993vy}.}
\begin{eqnarray}
H'&=&\Big[\frac{3-d-(d-1)r^2/\ell^2}{rV}-\frac{1}{r}\Big]H+\frac{\Omega}{2V}(H_++H_-)\nn\\
&&\label{Hsd}\\
H_{-}^{\prime}&=&\frac{{\cal
M}_d^2}{\Omega}H+\frac{d-2}{2r}H_++\Big[\frac{d-3+(d-1)
r^2/\ell^2}{2rV}-\frac{2d-3}{2r}\Big]H_-\label{Hsmd}
\end{eqnarray}
 with the constraint equation
\begin{eqnarray}
&&\hspace*{-2em}r^2\Omega\Big[4r\Omega^2-rV^{'2}+(d-2)VV^{'}+2rV{\cal
M}_d^2+2rVV^{''}\Big]H_--\Omega r^2V\Big[2{\cal
M}_d^2r+(d-2)V^{'}\Big]H_+\nn\\
&&\hspace*{12em}-2r^2V\Big[2(d-2)\Omega^2-2{\cal M}_d^2V+r{\cal
M}_d^2V^{'}\Big]H~=~0,\label{cons}
\end{eqnarray}
 where
\begin{eqnarray}
H\equiv H_{tr},~~~H_\pm\equiv\frac{H_{tt}}{V(r)}\pm
V(r)H_{rr}~~~{\rm
with}~~V(r)=1-\Big(\frac{r_0}{r}\Big)^{d-3}+\frac{r^2}{\ell^2}.
\end{eqnarray}
%%%%%%%%%%%%%%%%%%%%%%%%%%%%%%%%%%%%%%%%%%%%%%%%%%%%%%%%%%%%%%%%%%%%%%%%%%%%%%
%%%%%%%%%%%%%%%%%%%%%%%%%%%%%%%%%%%%%%%%%%%%%%%%%%%%%%%%%%%%%%%%%%%%%%%%%%%%%%
%%%%%%%%%%%%%%%%%%%%%%%%%%%%%%%%%%%%%%%%%%%%%%%%%%%%%%%%%%%%%%%%%%%%%%%%%%%%%%
\begin{figure*}[t!]
   \centering
   \includegraphics{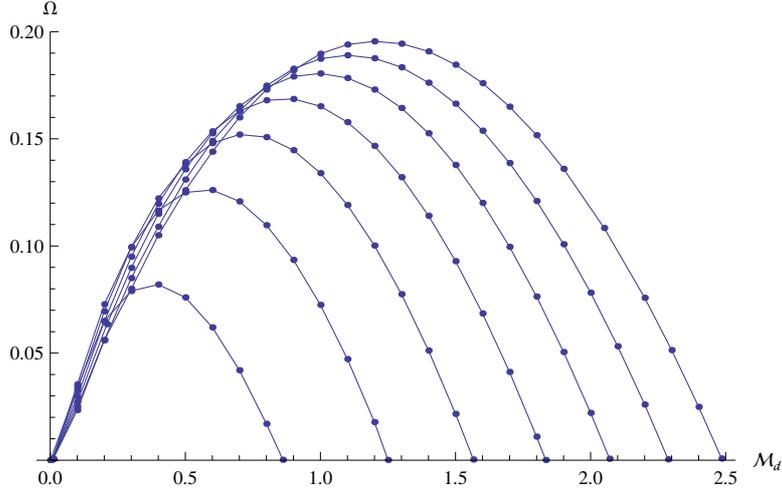}
\caption{$\Omega$ graphs as function of ${\cal M}_d$ for a small
black hole with $r_+=1,~\ell=10$ and $d=4,5,\cdots,10$ from left to
right curve. The most left curve in this figure corresponds to the
largest one in the figure 5.}
\end{figure*}
%%%%%%%%%%%%%%%%%%%%%%%%%%%%%%%%%%%%%%%%%%%%%%%%%%%%%%%%%%%%%%%%%%%%%%%%%%%%%%
%%%%%%%%%%%%%%%%%%%%%%%%%%%%%%%%%%%%%%%%%%%%%%%%%%%%%%%%%%%%%%%%%%%%%%%%%%%%%%
%%%%%%%%%%%%%%%%%%%%%%%%%%%%%%%%%%%%%%%%%%%%%%%%%%%%%%%%%%%%%%%%%%%%%%%%%%%%%%
At infinity of $r\to\infty$, asymptotic solutions to Eqs.(\ref{Hsd})
and ({\ref{Hsmd}) are
\begin{eqnarray}
H^{(\infty)}&=&C^{(\infty)}_{1}r^{-(d+1)/2+\sqrt{{\cal
M}_d^2\ell^2+(d-1)^2/4}} +C^{(\infty)}_{2}r^{-(d+1)/2-\sqrt{{\cal
M}_d^2\ell^2+(d-1)^2/4}},\nn\\
&&\nn\\
H_-^{(\infty)}&=&\tilde{C}^{(\infty)}_{1}r^{-(d-1)/2+\sqrt{{\cal
M}_d^2\ell^2+(d-1)^2/4}}
+\tilde{C}^{(\infty)}_{2}r^{-(d-1)/2-\sqrt{{\cal
M}_d^2\ell^2+(d-1)^2/4}},
\end{eqnarray}
where $\tilde{C}^{(\infty)}_{1,2}$ are
\begin{eqnarray}
\tilde{C}^{(\infty)}_{1}&=&\frac{{\cal M}_d^2}{(1-d)/2+\sqrt{{\cal
M}_d^2\ell^2+(d-1)^2/4}}{C}^{(\infty)}_{1}
,\nonumber\\
\tilde{C}^{(\infty)}_{2}&=& \frac{{\cal M}_d^2}{(1-d)/2-\sqrt{{\cal
M}_d^2\ell^2+(d-1)^2/4}}{C}^{(\infty)}_{2}.
\end{eqnarray}
 At the horizon
$r_+$, their asymptotic solutions are given by
\begin{eqnarray}
H^{(r_+)}&=&C^{(r_+)}_{1}(r^{d-3}-r_+^{d-3})^{-1+\Omega/V^{'}(r_+)}
+C^{(r_+)}_{2}(r^{d-3}-r_+^{d-3})^{-1-\Omega/V^{'}(r_+)},\nn\\
&&\nn\\
H_-^{(r_+)}&=&\tilde{C}^{(r_+)}_{1}(r^{d-3}-r_+^{d-3})^{\Omega/V^{'}(r_+)}
+\tilde{C}^{(r_+)}_{2}(r^{d-3}-r_+^{d-3})^{-\Omega/V^{'}(r_+)},
\end{eqnarray}
where $\tilde{C}^{(r_+)}_{1,2}$ are
\begin{eqnarray}
\tilde{C}^{(r_+)}_{1}&=&~\frac{(d-3)r_+^{d-3}\Omega\Big(2\Omega-V'(r_+)\Big)}{2V'(r_+)\Big({\cal
M}_d^2r_++(d-2)\Omega\Big)}~{C}^{(r_+)}_{1},\nn\\
\tilde{C}^{(r_+)}_{2}&=&~-\frac{(d-3)r_+^{d-3}\Omega\Big(2\Omega+V'(r_+)\Big)}{2V'(r_+)\Big({\cal
M}_d^2r_+-(d-2)\Omega\Big)}~{C}^{(r_+)}_{2}.\nn
\end{eqnarray}
We note that two boundary conditions of the regular solutions
correspond to ${C}^{(\infty)}_{1}=0$ and ${C}^{(r_+)}_{2}=0$ at
infinity and horizon, respectively.

Eliminating $H_+$ in Eqs. (\ref{Hsd}) and (\ref{Hsmd}) with the help
of the constraint (\ref{cons}), one can find the coupled equations
with $H$, $H_-$ only. For given dimensions $d=4,5,\cdots,10$, fixed
${\cal M}_d$, and various values of $\Omega$, we solve these coupled
equations numerically, which yields possible values of $\Omega$ as a
function of ${\cal M}_d$ given by
\begin{equation}
{\cal M}_d=\sqrt{\frac{dm^2_d}{4(d-1)}-\frac{(d-2)^2}{2\ell^2}}.
\end{equation}
Fig. 5 shows  that the curve of possible values of $\Omega$ and
${\cal M}_d$  intersects the ${\cal M}_d$-axis at two places: ${\cal
M}_d=0$  and ${\cal M}_d$ = ${\cal M}_d^c$ where  ${\cal M}_d^c$ is
a critical non-zero mass. The fact that the curve does not intersect
the  ${\cal M}_d$-axis at  ${\cal M}_d<0$ follows from the stability
of the AdS black hole in higher-dimensional Einstein gravity.
Explicitly, for ${\cal M}<{\cal M}_d^c({\cal M}>{\cal M}_d^c)$, the
AdS black hole is unstable (stable) against the metric
perturbations.  From the observation of Fig. 5, we read off the
critical mass ${\cal M}^c_d$ depending on the dimension $d$ as
\begin{equation} \label{cwn}
\left(
\begin{array}{c|ccccccc}
 d & 4 & 5 & 6 & 7 & 8 & 9 & 10 \cr \hline
      {\cal M}_d^c& 0.86 & 1.26 & 1.57 & 1.83 & 2.07 & 2.29& 2.49 \cr
\end{array}
\right).
\end{equation}
For higher-dimensional black strings with $V_{\rm
bs}(r)=1-(r_0/r)^{d-3}$, the critical wave number  marks the lower
bound of possible wavelengths for which there is an unstable mode.
Especially for $e^{\frac{\Omega}{r_0} t} e^{i\frac{k}{r_0} z}$
setting, there exists a critical  wave number $k^c_d$ where for
$k<k^c_d(k>k^c_d)$, the black string is unstable (stable) against
the metric perturbations. There is an unstable (stable) mode for any
wavelength larger (smaller) than the critical wavelength
$\lambda_{\rm GL}=2\pi r_0/k^c_d$: $\lambda>\lambda_{\rm
GL}(\lambda<\lambda_{\rm GL})$. The critical wave number $ k^c_d$
depends  on the dimension $d$ as~\cite{Harmark:2007md}
\begin{equation} \label{bwn}
\left(
\begin{array}{c|ccccccc}
 d & 4 & 5 & 6 & 7 & 8 & 9 & 10 \cr \hline
      k^c_d& 0.88 & 1.24 & 1.60 & 1.86 & 2.08 & 2.30& 2.50  \cr
\end{array}
\right).
\end{equation}
From $V(r)$ in (\ref{sch}), one has the relation between $r_+$ and
$r_0$ as \begin{equation}
r_0=\Big[\frac{r_+^{d-1}}{\ell^2}+r_+^{d-3}\Big]^{\frac{1}{d-3}}.\end{equation}
For $r_+=1$ and $\ell=10$, it takes the form
\begin{equation}
r_0=\Big(\frac{101}{100}\Big)^{\frac{1}{d-3}}
\end{equation}
which implies $r_0=\{1.01,1.005,1.003,1.002,1.002,1.002,1.001\}$.
Here, the corresponding $\tilde{k}_d^c=r_0{\cal M}^c_d$  is given by
\begin{equation} \label{kwn}
\left(
\begin{array}{c|ccccccc}
 d & 4 & 5 & 6 & 7 & 8 & 9 & 10 \cr \hline
      \tilde{k}^c_d& 0.87 & 1.27 & 1.57 & 1.83 & 2.07 & 2.29& 2.49 \cr
\end{array}
\right).
\end{equation}
W observe that $\tilde{k}_d^c=k_d^c- 0.01$ for $d=4,8,9,10$ whereas
$\tilde{k}_d^c=k_d^c\pm  0.03$ for $d=5,6,7$.

 Now let us derive the GL instability condition
from (\ref{kwn}). We propose the bound for unstable modes
approximately as
\begin{equation}
0<{\cal M}_d<\frac{\tilde{k}_d^c}{r_0}.
\end{equation}

We note  that there is no connection between thermodynamic
instability and classical stability for small AdS black hole in
higher-dimensional Einstein gravity. However, the GL instability
condition (massiveness) picks up the small AdS black hole with
$r_+<r^{(d)}_*$ which is thermodynamically unstable in fourth-order
gravity.  Hence, we conclude that there is a connection between the
GL instability and thermodynamic instability for small AdS black
hole in fourth-order gravity.

\section{Discussions}

First of all, we have studied the thermodynamics and phase
transitions of AdS black holes using the ADT thermodynamic
quantities in fourth-order gravity. For $m^2_d>m^2_{\rm c}$, all
thermodynamic properties are dominantly determined by Einstein
gravity, while for $m^2_d<m^2_{\rm c}$, all thermodynamic properties
are dominated by Weyl-squared term (conformal gravity). The former
is completely understood, but the latter has  a new feature when one
studies the black hole thermodynamics by using the ADT thermodynamic
quantities. A further study is necessary to understand  the latter
completely.

We have confirmed a close connection between thermodynamic and
classical  instability for the BTZ black hole in new massive
gravity.  Also, there is  a connection between the classical (GL)
 and thermodynamic instability for small AdS  black
holes in fourth-order gravity.   This implies  that the Gubser-Mitra
conjecture (CSC) holds for the AdS black holes found from
fourth-order gravity theory, which corresponds to our main result.

Finally, we wish to comment on the linearized equation
(\ref{lineqq}) for the metric perturbation, which is obtained by
splitting the linearized fourth-order equation. One confronts with
ghost states with negative kinetic term when one uses the
second-order equation (\ref{lineqq}) in fourth-order gravity. In
order to avoid this problem, one may express the linearized equation
(the linearized fourth-order equation for $h_{\mu\nu}$) in terms of
the linearized Einstein tensor as
\begin{equation} \label{glineqq}
\bar{\nabla}^2 \delta G_{\mu\nu}+2\bar{R}_{\alpha\mu\beta\nu}\delta
G^{\alpha\beta}-{\cal M}^2_d\delta G_{\mu\nu}=0
\end{equation}
which is surely a second-order differential equation.  This equation
describes  5 DOF of a massive spin-2 graviton propagating on the AdS
black hole spacetimes when one imposes the tracelessness of $ \delta
G^{\mu}~_\mu=-\delta R=0$ and the transversality  of
$\bar{\nabla}^\mu \delta G_{\mu\nu}=0$ from the contracted Bianchi
identity. Actually, Eq.(\ref{glineqq}) is a boosted-up version of
Eq.(\ref{lineqq}) which indicates the GL instability for small AdS
black holes. However, the former is a ghost free equation, while the
latter has  the ghost problem in fourth-order gravity.

\section*{Acknowledgement}

T.M. would like to thank  Sang-Heon Yi and Chong Oh Lee for useful
comments. This work was supported by the National Research
Foundation of Korea (NRF) grant funded by the Korea government
(MEST) (No.2012-R1A1A2A10040499). Y.M. was supported partly by the
National Research Foundation of Korea (NRF) grant funded by the
Korea government (MEST) through the Center for Quantum Spacetime
(CQUeST) of Sogang University with grant number 2005-0049409.

\newpage
\section*{Appendix: Coefficients of the perturbation equation for $H_{tr}$}

As was shown in Sec.4.2, the master equation for $H_{tr}$ is given
by
\begin{equation}\label{master} A(r;r_0,\ell,\Omega^2,{\cal M}_d^2)
\frac{d^2}{dr^2}H_{tr} +B\frac{d}{dr}H_{tr}+CH_{tr}=0,
\end{equation}
where $A,~B,$ and $C$ are

\begin{eqnarray}
A&=&-{\cal M}_d^2
V-\Omega^2+\frac{V^{'2}}{4}-\frac{VV^{''}}{2}-\frac{(d-2)VV^{'}}{2r},\nonumber\\
&&\nonumber\\
&&\nonumber\\
B&=&-2{\cal
M}_d^2V^{'}-\frac{3V^{'}V^{''}}{2}-\frac{3\Omega^2V^{'}}{V}+\frac{3V^{'3}}{4V}+\frac{(d-2){\cal
M}_d^2V}{r}+\frac{(d-2)\Omega^2}{r}+\frac{3(d-2)V^{'2}}{4r}\nonumber\\
&&\nonumber\\
&&+\frac{(d-2)VV^{''}}{2r}
-\frac{(d-2)^2VV^{'}}{2r^2},\nonumber\\
&&\nonumber\\
&&\nonumber\\
C&=&{\cal M}_d^4+\frac{\Omega^4}{V^2}+\frac{2{\cal
M}_d^2\Omega^2}{V}-\frac{5\Omega^2V^{'2}}{4V^2}+\frac{{\cal
M}_d^2V^{'2}}{4V}+\frac{V^{'4}}{4V^2}-\frac{{\cal
M}_d^2V^{''}}{2}-\frac{\Omega^2V^{''}}{2V}-\frac{V^{'2}V^{''}}{4V}-\frac{V^{''2}}{2}
\nonumber\\
&&\nonumber\\
&&-\frac{d{\cal
M}_d^2V^{'}}{2r}-\frac{(d-2)\Omega^2V^{'}}{2rV}+\frac{(d-2)V^{'3}}{2rV}-\frac{3(d-2)V^{'}V^{''}}{2r}
+\frac{(d-2)\Omega^2}{r^2}+\frac{(d-2){\cal
M}_d^2V}{r^2}\nonumber\\
&&\nonumber\\
&&-\frac{(d-2)(2d-3)V^{'2}}{4r^2}+\frac{(d-2)VV^{''}}{2r^2}+\frac{(d-2)^2VV^{'}}{2r^3}.
\nonumber
\end{eqnarray}
One can easily check that for $V=1-(r_0/r)^{d-3}$ or $V=1-r_0/r+
r^2/\ell^2$, Eq. (\ref{master}) reduces to the master equation for
$H_{tr}$ given in the literature \cite{Gregory:1993vy} or
\cite{Hirayama:2001bi}.

\end{document}